\documentclass[useAMS,a4paper]{mn2e}
\usepackage{graphicx}
\usepackage{amsmath}
\usepackage{amssymb}
\usepackage{mathptmx}

\title[A T\.ZO Candidate in the SMC]{Discovery of a Thorne-\.Zytkow object candidate in the Small Magellanic Cloud}
\author[E. M. Levesque et al.]{Emily M. Levesque$^{1,2}$\thanks{E-mail:
Emily.Levesque@colorado.edu}, Philip Massey$^{3}$, Anna N. \.Zytkow$^{4}$, and Nidia Morrell$^{5}$\\
$^{1}$Center for Astrophysics \& Space Astronomy, University of Colorado UCB 389, Boulder, CO 80309, USA\\
$^{2}$Hubble Fellow\\
$^{3}$Lowell Observatory, 1400 W. Mars Hill Road, Flagstaff, AZ 86001, USA\\
$^{4}$Institute of Astronomy, University of Cambridge, Madingley Road, Cambridge CB3 0HA, UK\\
$^{5}$Las Campanas Observatory, Carnegie Observatories, Casilla 601, La Serena, Chile}
\begin{document}

\date{Submitted 2014 April 17.}

\pagerange{\pageref{firstpage}--\pageref{lastpage}} \pubyear{2014}

\maketitle

\label{firstpage}

\begin{abstract}
Thorne-\.{Z}ytkow objects (T\.ZOs) are a theoretical class of star in which a compact neutron star is surrounded by a large, diffuse envelope. Supergiant T\.{Z}Os are predicted to be almost identical in appearance to red supergiants (RSGs). The best features that can be used at present to distinguish T\.{Z}Os from the general RSG population are the unusually strong heavy-element and Li lines present in their spectra, products of the star's fully convective envelope linking the photosphere with the extraordinarily hot burning region in the vicinity of the neutron star core. Here we present our discovery of a T\.{Z}O candidate in the Small Magellanic Cloud. It is the first star to display the distinctive chemical profile of anomalous element enhancements thought to be unique to T\.{Z}Os. The positive detection of a T\.{Z}O will provide the first direct evidence for a completely new model of stellar interiors, a theoretically predicted fate for massive binary systems, and never-before-seen nucleosynthesis processes that would offer a new channel for Li and heavy-element production in our universe.
\end{abstract}


\section{Introduction}
Thorne-\.{Z}ytkow objects (T\.{Z}Os) are a class of star originally proposed by Thorne \& \.{Z}ytkow (1975, 1977), comprised of a neutron star core surrounded by a large and diffuse envelope. Models for massive T\.{Z}Os (total mass $\gtrsim 11.5M_{\odot}$) predict that such a star would be primarily powered by thermonuclear reactions occurring at the base of the convective envelope, with gravitational accretion making only a small contribution ($\sim$5 per cent, e.g. Thorne \& \.{Z}ytkow 1977, Biehle 1991). T\.{Z}Os are expected to form as a result of the evolution of two massive stars in a close binary, with the neutron star forming when the more massive star explodes as a supernova. During subsequent evolution of the system, the expanding envelope of the companion may lead to a common envelope state and the spiral-in of the neutron star into the core of its companion (Taam et al.\ 1978). Alternately, Leonard et al.\ (1994) propose that a T\.{Z}O may be produced when a newly-formed neutron star receives a supernova `kick' velocity in the direction of its companion and becomes embedded.

T\.{Z}Os represent a completely new theoretical class of stellar object, offering a novel model for stellar interiors and an example of unique nucleosynthetic processes. However, there has never been a positive observational identification of a T\.{Z}O. The Thorne \& \.{Z}ytkow (1977) model predicts objects with an outward appearance that is virtually indistinguishable from normal M-type red supergiants (RSGs), lying at the Hayashi limit (Hayashi \& Hoshi 1961) and showing signs of excess mass loss. The only observational signature that can be used to identify T\.{Z}Os is an unusual chemical abundance pattern in their atmospheres. These anomalous element enhancements are the product of the `interrupted rapid-proton' process or {\it irp}-process, made possible by the extremely high temperatures at the surface of the neutron star core combined with the completely convective surrounding envelope (Cannon 1993; see also Wallace \& Woosley 1981). Although RSG spectra are dominated by broad absorption bands of TiO, several specific {\it irp}-process products should be observable as enhanced in a T\.{Z}O atmosphere, including lines of Rb I, Sr I and Sr II, Y II, Zr I, and Mo I (Biehle 1994). $^7$Li should also be over-abundant in T\.{Z}Os, a product of the $^7$Be-transport mechanism that is similarly dependent on the star's internal structure (e.g. Cameron 1955, Podsiadlowski et al.\ 1995). 

While several previous searches have been made for T\.{Z}Os, none have observed a star with an enhancement profile that is uniquely attributable to a T\.ZO interior - instead they measure only a single element enhancement such as Rb I (Kuchner et al.\ 2002) or an abundance profile that can be easily explained by known nucleosynthetic processes such as the $s$-process (Vanture et al.\ 1999). Until recently, another key difficulty in searching for T\.ZOs among the local RSG populations has been the lack of a large and well-characterised sample of RSGs from which to draw candidates and comparisons. Fortunately, in recent years updated atmosphere models and new observations have greatly improved our understanding of RSG samples in the Local Group (e.g. Levesque et al.\ 2005, 2006, 2007; Massey et al.\ 2009; Levesque \& Massey 2012). With this detailed picture of the nearby RSG population, we are now able to carefully select a sample of candidate T\.{Z}Os and use these to perform the first in-depth and extragalactic search for these stars. 

Here we present our observations of a T\.{Z}O candidate discovered during our search for T\.{Z}Os in the Milky Way and Magellanic Clouds. The Small Magellanic Cloud (SMC) star HV~2112 exhibits the distinctive set of anomalous element enhancements thought to be produced by processes characteristic to T\.{Z}Os and believed not to be associated with any other stars. We will discuss our sample selection and observation techniques (Section 2) as well as the analyses techniques used to search for atypical element abundances in these stars (Section 3). We discuss the physical properties of our T\.{Z}O candidate in detail (Section 4), and consider future work on these compelling objects (Section 5).

\section{Sample Selection and Observations}
We initially constructed our sample from our past effective temperature ($T_{\rm eff}$) studies of RSGs in the Milky Way and Magellanic Clouds (Levesque et al.\ 2005, 2006). We observed 24 RSGs in the Milky Way using the Astrophysics Research Consortium Echelle Spectrograph (ARCES; Wang et al.\ 2003) on the Apache Point Observatory 3.5-meter telescope on 11-12 Feb 2011 (UT).  We later extended this work to observations of 16 RSGs in the LMC and 22 in the SMC with the Magellan Inamori Kyocera Echelle (MIKE; Bernstein et al.\ 2003) and Magellan Echellete (MagE; Marshall et al.\ 2008) spectrographs on the Magellan 6.5-meter at Las Campanas Observatory on 13-15 Sep 2011. For our Magellanic Cloud observations we included several additional stars in our sample, selected based on 2MASS photometry and colours that were consistent with RSGs ($K < 8.9$, $J-K > 1$). 

Our ARCES observations of the Milky Way RSGs were performed using the default $1\farcs6 \times 3\farcs2$ slit. To achieve precise flatfield and wavelength calibrations for each star, we observed quartz lamps and ThAr lamps after each individual exposure. The spectra were reduced using standard IRAF procedures, using the ThAr lamp spectra for wavelength calibration. Each star's spectrum was corrected for radial velocity effects using the wavelengths of the Ca II triplet (8498.03\AA, 8542.09\AA, and 8662.14\AA).

Our Magellanic Cloud targets were observed with MIKE on 2011 Sep 13-15 and with MagE on 2011 Sep 14-16. For MIKE we used the $0\farcs7 \times 5\arcsec$ slit with 2$\times$2 binning, `slow' readout, and the standard grating settings, giving $R \sim 42,000$. With MagE we used the 1$"$ with 1$\times$1 binning and `slow' readout. Our observations were generally performed with seeing $\lesssim$1$"$ and airmasses $\lesssim$1.5. Our high-resolution MIKE spectra were observed for the purpose of measuring line ratios, while the MagE spectra were taken for the purpose of producing flux-calibrated spectrophotometry that could be used for model fitting and the determination of physical properties in the case of known variable stars and targets that had not previously been observed. The MagE spectra were taken at the parallactic angle to minimise light losses, and the spectrophotometric standards Feige 110, LTT 9491, LTT 1788, and LTT 2415 were observed for flux calibration (Oke 1990; Hamuy et al.\ 1992, 1999). Internal flats and ThAr lamps were observed for flatfielding and wavelength calibration purposes. The MIKE and MagE data were reduced using a combination of standard IRAF echelle routines and the \texttt{mtools} package (see Massey et al.\ 2012).

\section{Analysis of Element Enhancements}
High-resolution modeling of RSG atmospheres is subject to a number of physical complexities, including treatments of atmospheric geometry, optical depth variations, mass loss effects, and host galaxy abundance variations. To avoid potential biases introduced by the assumptions inherent to atmospheric modeling, we instead calculated equivalent width ratios for our spectral lines of interest. To search for signs of anomalous T\.ZO-like element enhancements in our spectra, we compared equivalent widths of Li, Rb, and Mo absorption features - all elements expected to be enhanced in T\.ZOs - to those of nearby spectral features where no significant enhancements are predicted such as K, Ca, Fe, and Ni (Cannon 1993, Biehle 1994, Podsiadlowski et al.\ 1995). 

In RSG spectra severe line blanketing effects completely obscure the continuum, making traditional measurements of absorption line equivalent widths impossible. We therefore adopt a method used by Kuchner et al.\ (2002) in their search for T\.{Z}Os, determining `pseudo-equivalenth widths' for the relevant lines. This technique uses the highest peaks in the surrounding region of the spectrum to define a `pseudo-continuum' within the IRAF \texttt{splot} function that can in turn be used to measure the integrated line strengths. In our analyses we take care to use the same definitions of the pseudo-continuum for each star in our sample. Both the spectral features driving the definition of the pseudo-continuum and the absorption features we are measuring will be dependent upon the star's $T_{\rm eff}$; as a result, all final analyses of the stars' observed line ratios are considered relative to $T_{\rm eff}$. Our measurements have systematic errors of $\lesssim$5 per cent.

We calculated the Rb I $\lambda$7800.23/Ni I $\lambda$7797.58 (Rb/Ni) line ratio and found good agreement between our sample and the values determined for Rb/Ni from Kuchner et al.\ (2002) using the same pseudo-equivalent width technique. We also measured the Rb I $\lambda$7800.23/Fe I $\lambda$7802.47 (Rb/Fe) and Mo I $\lambda$5570.40/Fe I $\lambda$5569.62 (Mo/Fe) ratios due to the proximity of these features in the spectrum; indeed, the Mo I and Fe I features are slightly blended. We used the Li I $\lambda$6707.97/K I $\lambda$7698.97 (Li/K) and Li I $\lambda$6707.97/Ca I $\lambda$6572.78 (Li/Ca) ratios to probe the relative lithium abundance in our stars. Li/K has previously been used to search for T\.{Z}O features in the supergiant U Aquarii (Vanture et al.\ 1999); the two features are both resonant lines (note that K I $\lambda$7698.97 is a blended doublet; Dunham 1937) and have similar ionisation potentials ($\chi_{\rm ion}$ = 5.4 eV for Li I and $\chi_{\rm ion}$ = 4.3 eV for K I). Similarly, Li/Ca has been previously used to examine lithium abundance in the supergiant S Persei (Gahm \& Hultqvist 1976), citing the lines' common ground state and photospheric origin. Finally, to ensure that our ratios were true measures of abundance anomalies, and not dominated by uncertainties due to weak comparison features, we also calculated line ratios for Ni I $\lambda$7797.58/Fe I $\lambda$7802.47 (Ni/Fe), K I $\lambda$7698.97/Ca I $\lambda$6572.78 (K/Ca), and Ca I $\lambda$6572.78/Fe I $\lambda$65569.62 (Ca/Fe) for our stars.

We compared the line ratios for our Milky Way and Magellanic Cloud stars relative to the stars' $T_{\rm eff}$. To define what would be considered an atypically enhanced line ratio, we determined the best linear fit for each set of ratios within a galaxy as a function of $T_{\rm eff}$, and calculated the 3$\sigma$ variations from these fits that should encompass 99.7 per cent of a normally-distributed sample. A ratio is considered statistically anomalous if it falls outside the 3$\sigma$ range.

\section{The T\.{Z}O Candidate HV~2112}
When comparing the ratios of T\.{Z}O-enhanced elements for the RSGs in each host galaxy, we found that HV~2112 ($\alpha_{\rm 2000}$=01:10:03.87, $\delta_{\rm 2000}$=$-$72:36:52.6), a member of our SMC sample, has anomalously high ($>$3$\sigma$ deviation from the mean) values of Rb/Ni, Li/K, Li/Ca, and Mo/Fe. The ratios measured for our full SMC sample, including HV~2112, are shown in Figure 1. The high ratios of HV~2112, combined with typical ratios for Ni/Fe and K/Ca, present clear evidence of Rb, Li, {\it and} Mo enhancement in the star's atmosphere, a combination thought to only be possible as a result of the exotic stellar interiors of T\.ZOs. A comparison of the HV~2112 spectrum to a `typical' SMC RSG is shown in Figure 2; the HV~2112 spectrum exhibits notably stronger T\.{Z}O features. 

While Rb and Mo have been individually observed in stars where their presence has been attributed to the $s$-process, there are no observed or predicted examples of the $s$-process producing both in a single star along with an additional Li enhancement. The presence of Li thus provides an additional argument; simultaneous Rb and Li enhancement is not observed in cool $s$-process stars (e.g. Garc\'{i}a-Hern\'{a}ndez et al.\ 2013), and the combination of these three elements has never been observed in any $s$-process stars (or, indeed, any other cool massive star). We also consider the strength of the Ba II 4554\AA\ absorption feature; since Ba is a common $s$-process product this offered an additional means of determining whether enhancements in HV~2112 could be attributed to the $s$-process rather than the {\it irp}-process (see Vanture et al.\ 1999). The strength of Ba II 4554\AA\ in HV~2112 did not show any signs of $s$-process enhancement compared to the rest of the SMC RSG sample.

\begin{figure}
\includegraphics[width=\columnwidth]{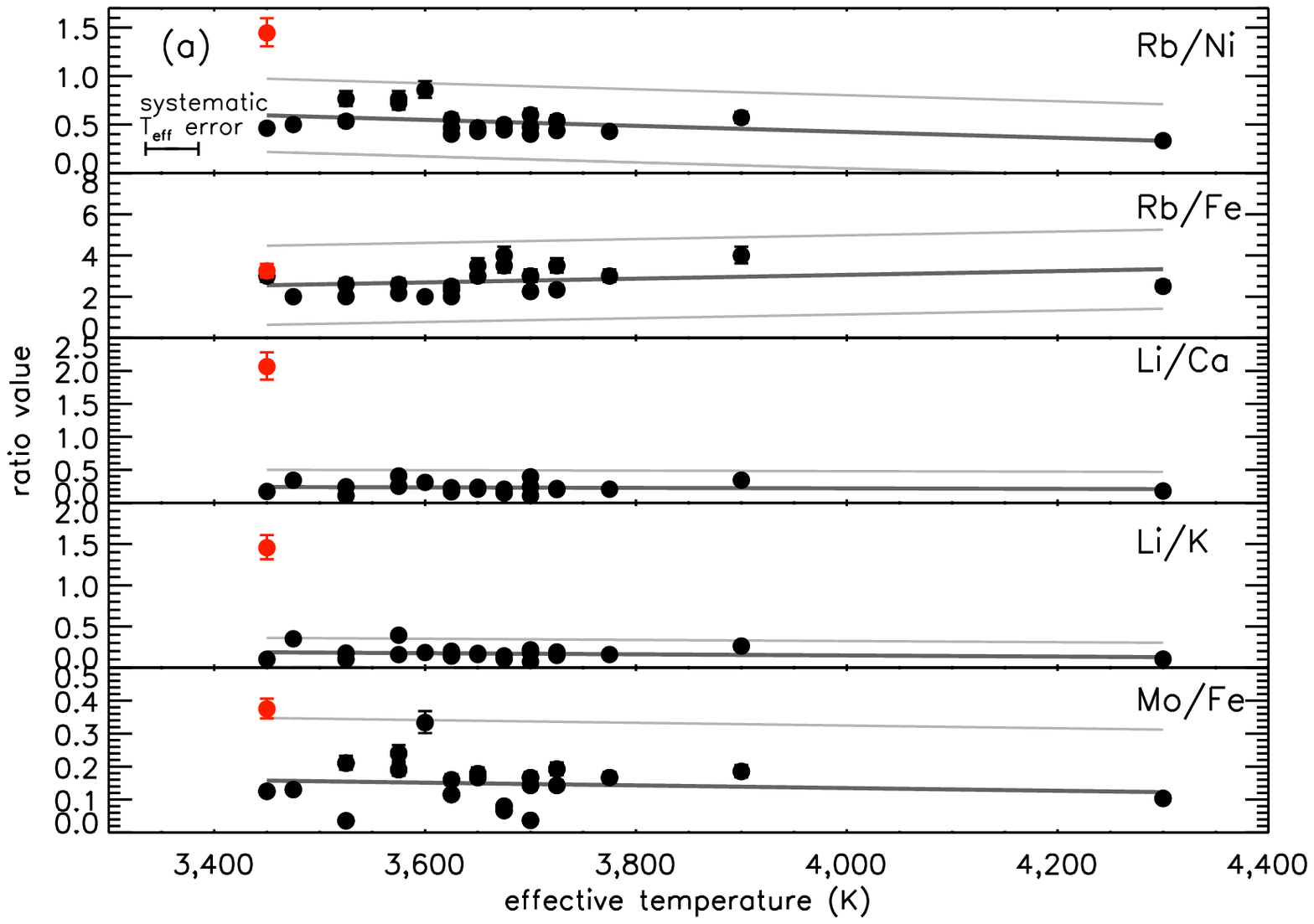}
\includegraphics[width=\columnwidth]{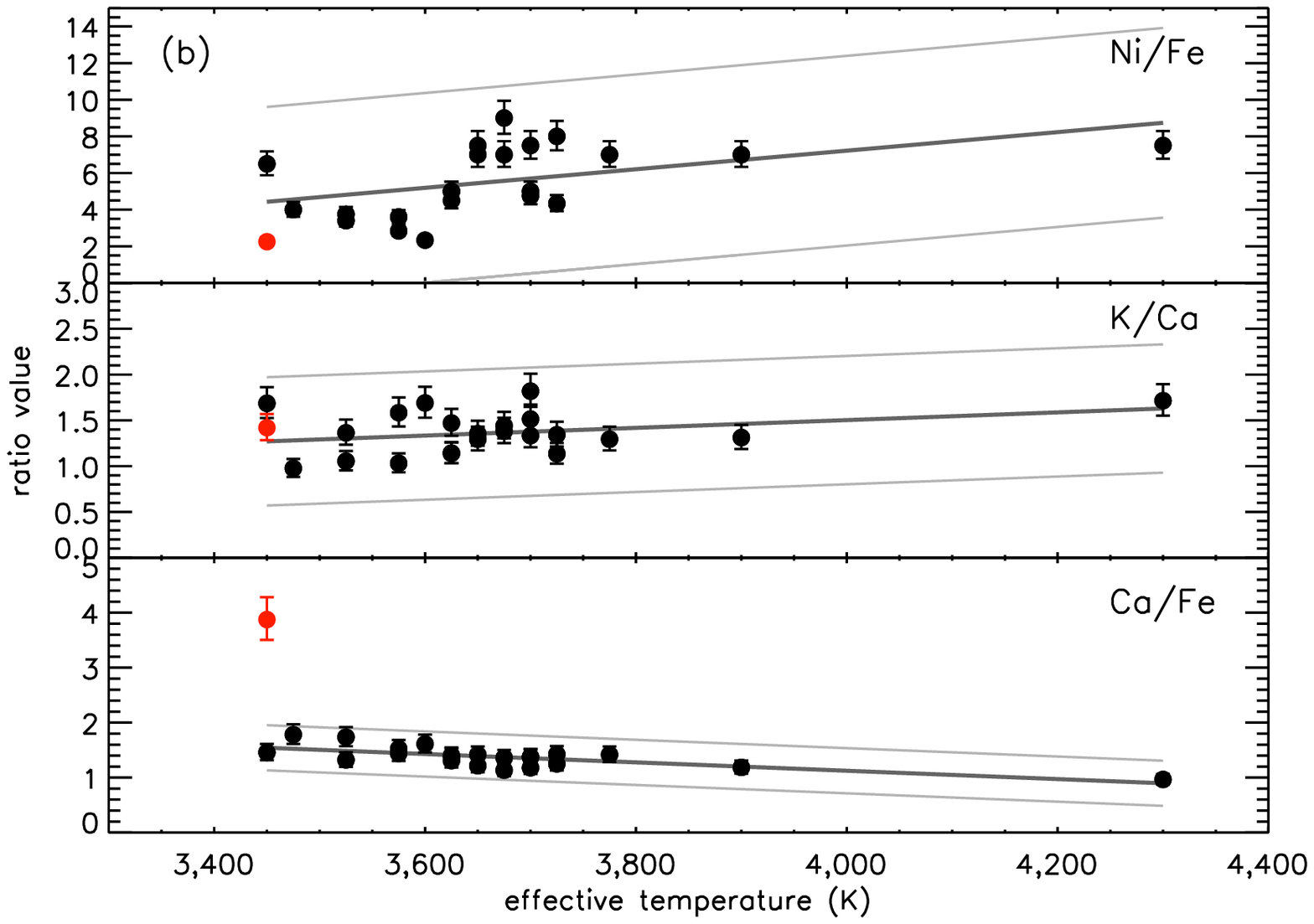}
\caption{ $T_{\rm eff}$ vs. measured line ratios for our sample of SMC targets. The ratios include features that we expect to be enhanced in T\.{Z}Os (a) as well as our `control' features (b). A dark grey line shows the best linear fit for each line ratio as a function of $T_{\rm eff}$, while the light grey lines mark the 3$\sigma$ deviations from the fit that should encompass 99.7 per cent of a normally-distributed sample. Ratios measured in our spectrum of HV~2112 are plotted in red. Error bars for each point illustrate the systematic errors of $\lesssim$5 per cent in our equivalent width measurements.}
\label{090426:HR}
\end{figure}
\begin{figure}
\includegraphics[width=\columnwidth]{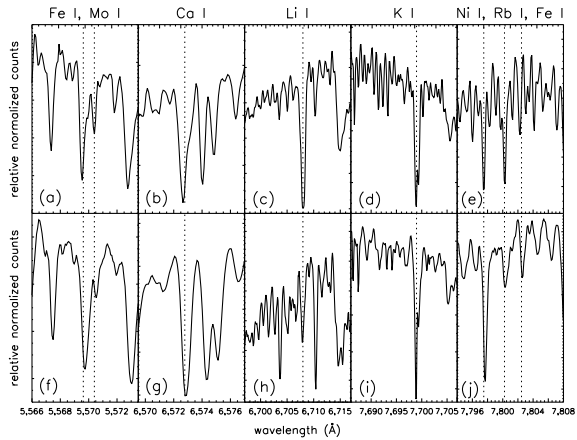}
\caption{Spectral features of HV 2112 and [M2002] SMC 005092 used in our analyses. [M2002] SMC 005092 (f-j) represents a `typical' SMC RSG with line ratios in agreement with the rest of our sample and a $T_{\rm eff}$ of 3475 K, comparable to the 3450 K $T_{\rm eff}$ we measure for HV~2112 (a-e). The plotted features include Fe I $\lambda$5569.62 and Mo I $\lambda$5570.40 (a,f), Ca I $\lambda$6572.78 (b,g), Li I $\lambda$6707.97 (c,h), K I $\lambda$7698.97 (d,i), and Ni I $\lambda$7797.58, Rb I $\lambda$7800.23, and Fe I $\lambda$7802.47 (e,j). Spectra are from our MIKE observations; both have a median S/N $\sim$ 10 and have been corrected to rest-frame. Wavelengths of the absorption features of interest are marked as vertical dashed lines.}
\end{figure}

\subsection{Physical Properties of HV~2112}
\subsubsection{Past Work}
HV~2112 is strongly photometrically variable, with the Harvard Variable catalog reporting a $\Delta$m of 4.80 mag and a $\sim$600d period (Payne-Gaposchkin \& Gaposchkin 1966; see also Reid \& Mould 1990). It is also a spectroscopic variable; previous studies have assigned it spectral types ranging from M3e to M7.5 (Wood et al.\ 1983, Reid \& Mould 1990, Smith et al.\ 1995), similar to other spectroscopically-variable RSGs (Levesque et al.\ 2007). HV~2112 was not identified as a T\.{Z}O candidate by past studies due to observational limitations; these prior spectra of the star lacked sufficient resolution (Wood et al.\ 1983) or wavelength coverage (Smith et al.\ 1995) for examining HV 2112's {\it irp}-process abundances. 

HV~2112 was previously classified as an asymptotic giant branch (AGB) star rather than a RSG (Wood et al.\ 1983, Reid \& Mould 1990, Smith et al.\ 1995). However, there are two problems with this classification. First, Wood et al.\ (1983) determined a bolometric magnitude ($M_{\rm bol}$) of $-7.22$ mag based on applying a mean $K$-band bolometric correction (BC) of 2.85 mag to the observed $K$ magnitude. The use of a mean BC can lead to significant errors for very hot or cool stars, as the BC is a steep function of $T_{\rm eff}$ even in $K$. More recent work has determined $T_{\rm eff}$-dependent fits for RSG bolometric corrections in multiple bands (Levesque et al.\ 2005, 2006); based on that work we determine a $M_{\rm bol}$ of $-7.8\pm0.2$ for HV~2112 (see below). Second, even the $M_{\rm bol} = -7.22$ derived by Wood et al.\ (1983) places the star as slightly more luminous than the Paczy\'{n}ski (1971) AGB upper luminosity limit of $-7.1$, and hence in better agreement with the luminosity range expected for RSGs.  Finally, for a known variable such as HV~2112 it is critical that contemporaneous measurements of $T_{\rm eff}$ and magnitude are used when determining $M_{\rm bol}$ and applying a bolometric correction.

\subsubsection{This Work}
From fitting a MagE spectrum of HV~2112 taken the night before our MIKE observations with the MARCS stellar atmosphere models (Gustafsson et al.\ 2008), we find the HV~2112 has a $T_{\rm eff}$ of 3450 K and a spectral type of M3 I, along with a $V$ magnitude of $13.7\pm0.1$ (see Figure 3). From our observations of the Ca II triplet in the HV~2112 spectrum we measure a radial velocity (RV) of $\sim$157 km s$^{-1}$ for this star. This is consistent with it being a true member of the SMC rather than a foreground dwarf (Neugent et al.\ 2010), even when considering the potential contribution from a supernova kick velocity of up to $\pm$75 km s$^{-1}$ (the average runaway velocity of a collisionally-formed T\.{Z}O; Leonard et al.\ 1994).  To determine $M_{\rm bol}$ we corrected the star's $V$ magnitude for the SMC distance modulus of 18.9 (van den Bergh 2000) and the $T_{\rm eff}$-dependent BC in $V$ (Levesque et al.\ 2006). With these contemporaneous values of $V$ and $T_{\rm eff}$, we calculated $M_{\rm bol}=-7.82\pm0.2$. This corresponds to a RSG with log $g = 0.0$ and an initial mass of $M\sim15M_{\odot}$ (Maeder \& Meynet 2001). This is well above the maximum mass limit for even super-AGB stars (e.g. Siess 2006, Poelarends et al.\ 2008), confirming that HV~2112 is indeed a RSG.
\begin{figure}
\includegraphics[width=\columnwidth]{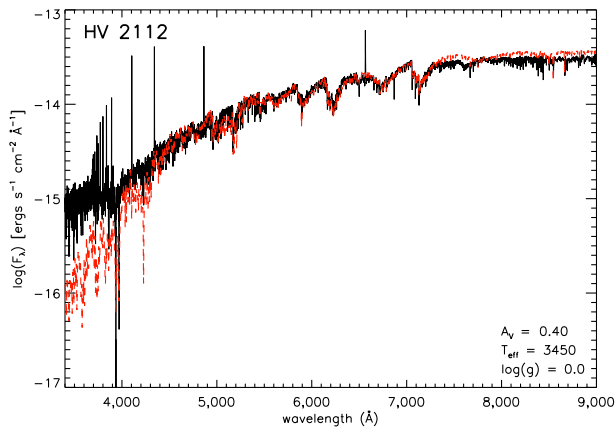}
\caption{Spectral energy distribution of HV~2112. Our MagE spectrum of HV~2112 is plotted (solid black line) along with the best-fit MARCS stellar atmosphere model (dashed red). The strong Balmer emission features in the HV~2112 spectrum are clearly visible, along with the near-UV excess and near-IR deficiency typical of circumstellar dust effects in RSG spectra (Levesque et al.\ 2005, Massey et al.\ 2005)}
\end{figure}

We observe an excess amount of visual extinction in the direction of HV~2112 ($A_V\sim0.4$ as compared to the average $A_V=0.24$ for SMC OB stars assuming a standard $R_V = 3.1$; Cardelli et al.\ 1989, Massey et al.\ 1995) and a slight flux excess ($\lesssim$10 ergs s$^{-1}$ cm$^{-2}$ \AA$^{-1}$) in the near-UV region of the spectral energy distribution. Both of these features have previously been observed in a number of other Milky Way and Magellanic Cloud RSGs, and are thought to be signatures of excess circumstellar dust associated with strong mass loss (see Levesque et al.\ 2005, Massey et al.\ 2005). HV~2112 is also spectroscopically variable and lies beyond the Hayashi limit for SMC supergiants (see Levesque et al.\ 2006, 2007); all of this is consistent with predictions for T\.ZOs (Thorne \& \.Zytkow 1977, van Paradijs 1995). Finally, closely-orbiting massive binaries are more common at low metallicities (Linden et al.\ 2010), making the predicted binary progenitor scenarios for T\.ZOs more likely in metal-poor host environments such as the SMC (log(O/H) + 12= 8.0, van den Bergh 2000).

\subsection{Additional Spectral Features}
HV~2112 exhibits several unusual spectral features not previously predicted to be associated with T\.{Z}Os. Most notable are the strong hydrogen Balmer emission features apparent in the star's spectrum (which is otherwise consistent with a cool mass-losing RSG; see Figure 3), extending from H$\alpha$ out to H18 (3691.55\AA). It is the only star in our sample to show such emission. Although HV~2112 has previous been classified as an occasiona; Me-type star (Wood et al.\ 1983), its strong Balmer line emission spectrum has not been previously presented or discussed. 

Some RSGs do show Balmer emission along with other features typical of HII region emission, which has been attributed to an ionised circumstellar environment (e.g. Levesque et al.\ 2009, Wright et al.\ 2014) Alternately, RSGs with hot OB star companions also show Balmer emission in their spectra (Cowley 1969). However, the spectrum of HV~2112 does {\it not} include any other emission features typical of an ionised nebula (such as [NII], [OII], or [OIII]). The spectral energy distribution also shows no sign of a hot companion capable of producing these strong hydrogen lines. If a hot companion was present, we would expect poor agreement between the model and the star's spectral energy distribution except over very short wavelength ranges (the slight excess at $<4000\AA$ in Figure 3 is a common signature of circumstellar extinction, as discussed above).

The Balmer emission spectrum of HV~2112 is most similar to that observed in M-type Mira variables (albeit with considerably stronger emission features throughout), showing a Balmer {\it increment} with H$\delta$ as the strongest feature followed by H$\gamma$, H$\beta$, and H$\alpha$ as well as the higher-order Balmer lines (see Figure 3). This inverted Balmer emission in M-type Mira variables is thought to be produced by non-LTE radiative transfer effects in the hydrogen lines from formation in a shocked atmosphere (Castalez et al.\ 2000), with shocks generated by pulsation and propagating through the atmosphere. As a supergiant, HV~2112 cannot be included as a member of the Mira variable class of giant stars; however it does show similar photometric and spectroscopic variability to these stars. It is possible that it could also produce the pulsationally-driven shocks that would give rise to this Balmer emission profile.

In addition, several of the line ratios observed in HV~2112 are unexpected. While the Rb/Ni ratio in HV~2112 is far higher than the average ratio measured for SMC RSGs, the Rb/Fe ratio is quite typical. The Ni/Fe ratio in HV~2112 is also typical of the RSG sample, precluding the possible explanation of a Fe overabundance. The Mo, Li, and Rb features do not, at a glance, exhibit the extremely increased strengths one might expect based on the large abundance enhancements predicted for T\.{Z}Os (Cannon 1993, Biehle 1994, Podsiadlowski et al.\ 1995), although it is important to recall that abundance-driven changes in line strengths are dependent on the curve of growth. Relatively weak enhancements could be indicative of an early or short-lived T\.{Z}O phase (Podsiadlowski et al.\ 1995), as predicted by theoretical considerations (Thorne \& \.{Z}ytkow 1975, 177). The Ca/Fe ratio in HV~2112 is also anomalously high, an enhancement not currently associated with T\.{Z}Os. However, it is important to note that convection models for massive star envelopes (particularly at low temperatures) have advanced significantly in recent years. As a result, the surface enhancements that would be expected for T\.{Z}Os could be potentially altered. This suggests that further work on the predicted observable abundances in T\.{Z}Os is needed, as already indicated in previous studies (Cannon 1993, Biehle 1994).

Finally, while the kinematics of HV~2112 are consistent with SMC membership, we cannot rule out the possibility that it may be a halo giant with a similar radial velocity. Classification as a Milky Way giant would eliminate the possibility that HV~2112 is a T\.{Z}O supergiant; however, the element enhancements and emission features that we observe would still require a novel explanation.

\section{Conclusions and Future Work}
Our observations of HV~2112 represent the most encouraging detection of a T\.{Z}O candidate to date. The anomalous enhancements of the Rb/Ni, Li/K, Li/Ca, and Mo/Fe ratios in HV~2112 satisfy the basic criteria for T\.{Z}O detection put forth by previous surveys and models (e.g. Biehle 1994, Vanture et al.\ 1999, Kuchner et al.\ 2002, Podsiadlowski et al.\ 1995): so far the only explanations for this chemical signature are the nucleosynthetic processes unique to the internal structure of a T\.{Z}O. At the same time, HV~2112 exhibits several additional oddities in its spectrum that are not predicted by current T\.{Z}O models.

Theoretical modeling of convective stellar envelopes has undergone major advances in recent years, with improved treatments of mixing length theory, overshooting, and more extensive reaction networks. As a result, new models of T\.ZOs will provide more realistic predictions of element enhancements, and may also predict additional identifying observables. This new theoretical work is the clear next step in testing whether HV~2112 is indeed a bona fide T\.ZO. Definitive detection of a T\.{Z}O would provide direct evidence for a completely new model of stellar interiors, as well as confirm a theoretically predicted fate for massive star binary systems and the existence of nucleosynthesis environments that offer a new channel for heavy-element and lithium production in our universe. As a compelling candidate, HV~2112 will serve as a valuable archetype in future searches for these objects.

\section*{Acknowledgements}
This paper is based on data gathered with the 6.5m Magellan telescopes located at Las Campanas, Chile and the Apache Point Observatory 3.5-meter telescope, which is owned and operated by the Astrophysical Research Consortium. EML is supported by NASA through Hubble Fellowship grant  number HST-HF-51324.01-A from the Space Telescope Science Institute, which is operated by the Association of Universities for Research in Astronomy, Incorporated, under NASA contract NAS5-26555. PM acknowledges support from NSF grant AST-1008020. AN\.{Z} thanks the Mitchell Family Foundation for support as well as Texas A\&M University and Cook's Branch Nature Conservancy for their hospitality.

\footnotesize{

}

\label{lastpage}
\end{document}